\documentclass[conference]{IEEEtran}
\IEEEoverridecommandlockouts
\usepackage{multirow}
\usepackage{graphicx}
\usepackage[style=ieee]{biblatex}
\usepackage{amsmath,amssymb,amsfonts}
\usepackage{algorithmic}
\usepackage{graphicx}
\usepackage{textcomp}
\usepackage{xcolor}
\usepackage{tabularx}
\usepackage[hidelinks]{hyperref}

\def\BibTeX{{\rm B\kern-.05em{\sc i\kern-.025em b}\kern-.08em
    T\kern-.1667em\lower.7ex\hbox{E}\kern-.125emX}}
\begin{document}

\title{Applying CodeBERT for Automated Program Repair of Java Simple Bugs\\
}

\author{\IEEEauthorblockN{Ehsan Mashhadi}
\IEEEauthorblockA{\textit{Schulich School of Engineering} \\
\textit{University of Calgary}\\
Calgary, Canada \\
ehsan.mashhadi@ucalgary.ca}
\and
\IEEEauthorblockN{Hadi Hemmati}
\IEEEauthorblockA{\textit{Schulich School of Engineering} \\
\textit{University of Calgary}\\
Calgary, Canada \\
hadi.hemmati@ucalgary.ca}
}

\maketitle

\begin{abstract}
Software debugging, and program repair are among the most time-consuming and labor-intensive tasks in software engineering that would benefit a lot from automation. In this paper, we propose a novel automated program repair approach based on CodeBERT, which is a transformer-based neural architecture pre-trained on large corpus of source code. We fine-tune our model on the ManySStuBs4J small and large datasets to automatically generate the fix codes. The results show that our technique accurately predicts the fixed codes implemented by the developers in 19-72\% of the cases, depending on the type of datasets, in less than a second per bug. We also observe that our method can generate varied-length fixes (short and long) and can fix different types of bugs, even if only a few instances of those types of bugs exists in the training dataset. 
\end{abstract}


\section{Introduction}
The goal of automated program repair (APR) techniques is to change an existing buggy program, automatically, to fix its bug(s), which has been investigated by many researchers in recent years \cite{monperrus2018automatic}. Given the similarity between a program repair task and generic natural language processing (NLP) tasks such as sequence to sequence learning and machine translation, in recent years, there has been a lot of work on applying machine learning for program repair \cite{tufano2018empirical}, \cite{chen2019sequencer}, \cite{li2020dlfix}, and \cite{lutellier2020coconut}. 

Within the field of NLP, one of the recent success stories is the use of large-scale pre-trained language models such as BERT \cite{devlin2018bert}. The advantages of using a pre-trained model are to leverage the very large-scale training set and fine-tune the model for the particular task in hand. However, BERT was only trained on natural language corpus.  More recently, Microsoft Research has released the CodeBERT model, which is a bimodal pre-trained language model for both natural and programming languages \cite{feng2020codebert}. This model is trained with a dataset provided by Husain et al. \cite{husain2019codesearchnet} that includes 6.4M unimodal codes in different programming languages such as Java, Python, Go, JavaScript, PHP, and Ruby. They trained CodeBERT with two objectives, masked language modeling (MLM) and replaced token detection (RTD). Reported results on the application of CodeBERT for a documentation generation task on the CodeSearchNet Corpus \cite{husain2019codesearchnet} show that it outperforms all baselines.

Given the promising results on similar tasks, in this paper, we leveraged CodeBERT to automatically generate fixes for bugs reported on the ManySStuBs4J dataset \cite{karampatsis2020often}. The ManySStuBs4J dataset focuses on Java simple bugs that appear on a single statement and the corresponding fix is within that statement.
To guide our investigation, we target answering the following research questions:

\textbf{RQ1. Can CodeBERT be used to fix Java simple bugs, and what are the pros and cons?} We found that our approach has an accuracy of 72\% and 68.8\% for the large and small versions of the dataset, respectively. Also, we observed that the accuracy is reduced to 23.7\% and 19.65\%, respectively, for unique datasets (after removing duplicate fixes from the datasets). Our approach does not require any special tokens for locating bugs such as SequenceR \cite{chen2019sequencer}, nor needs context lines, such as what some previous work (DLFix \cite{li2020dlfix} and CoCoNuT \cite{lutellier2020coconut}) require.  

Another pros of our technique is that it works well with small datasets as well as large ones. Given that CodeBERT is pre-trained, we just need to fine-tune it on the local training set, but previous works usually need a large dataset to train their models from scratch, which is an expensive and time-consuming task. Finally, our technique can be easily applied to other datasets and programming languages because it does not need any language-specific lexer or parser.

\textbf{RQ2. What are the characteristics of the fixed bugs by CodeBERT?} We investigated the characteristics of bugs fixed by our approach to find their effects on our technique. We found that our model is able to generate patches for different bug types even if only a few instances of those types exist in the local training dataset. Also, we realized that our technique is efficient in generating fix codes whose elements/tokens are not available in the local training dataset. Our analysis showed that common bugs in Java projects such as using a wrong Boolean parameter or passing parameters to a function with the wrong order are perfectly fixed with our technique. Finally, the varied-length automated fixes indicate that our approach can generate long patches as well as small patches.

\section{Related Work}
Most of the earlier APR works leverage search-based software engineering to generate patches. They mutate the buggy code by using some operators and they apply a search strategy to find the fixed code. For example, GenProg \cite{le2011genprog} tries to repair programs by leveraging genetic search at the statement level by using codes of the same program. MutRepair \cite{martinez2016astor} generates patches by using mutation operators to fix the bugs within if-condition statements. Our approach is different from these works since we do not use search-based techniques.

There are also some works specific to compilation errors. DeepFix \cite{gupta2017deepfix} is an end-to-end solution that tries to fix common C language multi-line compiler errors by using a neural network. TRACER \cite{ahmed2018compilation} is another tool for fixing compiler errors by focusing on single-line errors. These works focus on fixing the compilation errors which are easy to be verified by compiling the code or using the language-specific parser, but our technique can generate patches for logical errors.

Some other works mine bugs to learn fixing patterns from past bug fixes. For example, HDRepair \cite{le2016history} repairs bugs by mining recurrent patterns from real bug fixes. Prophet \cite{long2016automatic} learns from previously successful human patches, and Genesis \cite{long2017automatic} generates patches automatically from previously submitted patches. DeepRepair \cite{white2019sorting} leverages machine learning techniques to select similar code by leveraging recursive autoencoders to find repair ingredients from codes that are similar to the buggy code. DLFix \cite{li2020dlfix} shows that these pattern-based approaches have lower accuracy than the DL-based techniques on the popular Defects4j \cite{just2014defects4j} dataset.

Recent works try to use machine translation technique to fix the bugs which are more similar to our approach. Sequencer \cite{chen2019sequencer} uses seq2seq learning by combining an encoder/decoder architecture. The encoder and decoder are recurrent neural networks using LSTM gates. It overcomes the problem of large vocabulary in source code by leveraging the copy mechanism. Also, it performs a Buggy Context Abstraction process to organize the fault localization data and overcome the dependency problem. It needs special tokens such as \textless START\_BUG\textgreater  and \textless END\_BUG\textgreater for indicating the start and end of bugs. 

Ratchet \cite{hata2018learning} leverages the seq2seq translation to generate patches by using NMT with attention-based Encoder-Decoder. Tufano et al. \cite{tufano2018empirical} investigate the potential of NMT to generate candidate patches by using a recurrent neural network (RNN) encoder-decoder. They use simple code abstractions to make the source code smaller, which relies on a Lexer and Parser. CoCoNuT \cite{lutellier2020coconut} is an end-to-end approach using NMT and ensemble learning to automatically repair bugs in multiple languages. They used CNNs instead of RNNs used by previous works such as SeuqenceR \cite{chen2019sequencer} and Tufano et al. \cite{tufano2018empirical}. CoCoNuT \cite{lutellier2020coconut} represents the buggy source code and its surrounding context separately to improve the results. The authors evaluate this method on six benchmarks for four programming languages including Java, C, Python, and JavaScript. DLFix \cite{li2020dlfix}, is a two-tier DL model that leverages prior fixes and the surrounding code contexts to fix the new buggy code. The main difference between our work and these approaches are not requiring special tokens like SequenceR \cite{chen2019sequencer}, and there is no need for separating inputs of buggy line and context like \cite{lutellier2020coconut}. Also, we did not leverage any abstraction process like Tufano et al. \cite{tufano2018empirical}.

\section{Experiment}
In this section, we explain our datasets, experiment design, and results. 

\subsection{Dataset}
The ManySStuBs4J dataset has small and large versions consist of 10,231 and 63,923 instances, respectively. These instances are single statement bugs mined from 12,598 and 86,771 bug-fix commits, respectively, with only single-statement changes. For the purpose of our work, we consider a subset of this data named ``Duplicate'' because some instances with the type of \verb#Missing Throws Exception#, \verb#Delete Throws Exception#, and \verb#Change Modifier# do not contain \verb#sourceBeforeFix# and \verb#sourceAfterFix#, properly. 

We also created a new ``Unique'' dataset by removing duplicate \textless buggy code, fixed code\textgreater  pairs from both large and small datasets. The original datasets and our datasets statistics are presented in Table \ref{tab_dataset}. The table shows that we considered more than 91\% and 80\% of original large and small datasets, respectively. We split our datasets randomly into training, validation, and test datasets by using the 80\%, 10\%, 10\% ratio. 

\subsection{Design}
The CodeBERT follows BERT\cite{devlin2018bert} and RoBERTa \cite{liu2019roberta} to use a multi-layer bidirectional Transformer as the architecture. Its architecture is the same as RoBERTa-base, and the input format for pre-training step is concatenation of two segments with a separator token that is $[CLS], w\textsubscript{1},w\textsubscript{2},…,w\textsubscript{n},[SEP],c\textsubscript{1},c\textsubscript{2},...,c\textsubscript{m},[EOS]$. The first part contains natural language (NL) text, and the second part contains code (PL). The output includes contextual vector representation of NL and PL, and $[CLS]$ representation is used as the aggregated sequence representation for ranking or classification purposes. The authors trained CodeBERT by setting the maximum sequence length to 512 tokens, but since CodeBERT needs two extra tokens internally, so it is not possible to feed codes with token lengths large than 510, directly. We found that this is not a serious problem in our case since the duplicate and unique dataset contains only 173 instances out of 66,461 instances and 122 instances out of 29,168 instances, respectively, where the code token length is greater than 510. It is however possible that CodeBERT has seen parts of the test set during pre-training, but since its objective is not a program repair task,  so it did not try to find the relationship between fixed code from buggy code. The authors proposed a downstream task for generating code documentation named code2nl to evaluate their model's effectiveness in generation tasks.

\begin{table}[htbp]
\caption{The Numbers of Buggy and Fixed Code Pairs Per Dataset}
\begin{center}
\begin{tabular}{|c|l|l|l|l|l|l|}
\hline
\textbf{Name}&
\multicolumn{2}{|c|}{Original}&
\multicolumn{2}{|c|}{Duplicate}& \multicolumn{2}{|c|}{Unique}\\
\hline
\textbf{Size}&Large&Small&Large&Small        &Large&Small\\
\hline
\textbf{Number}&63,923&10,231&58,198&8,263&24,488&4,680\\
\hline
\end{tabular}
\label{tab_dataset}
\end{center}
\vspace*{-10pt}
\end{table}

Since the CodeBERT is just an encoder, so they used a transformer decoder attached to the encoder part to make this model usable for generation tasks like document generation. They compared their approach with baselines by calculating the BLEU score of generated documentation which confirms its superiority. The BLEU score is more like a similarity metric, which is quick, inexpensive, and language-independent, and correlates highly with human evaluation \cite{papineni2002bleu}. They used a Transformer with six layers, 768-dimensional hidden states, and 12 attention heads as a decoder. Also, the Adam optimizer is used for updating the model parameter in conjunction with performing early stopping on the validation set.

We leveraged this downstream task for fixing program bugs, and since our approach does not need any natural language texts, we just provide code to the model by using the second part of the input format which is after $[SEP]$ token. For our task in hand, we used buggy codes as a source side and the fixed codes as the target side. We set the learning rate to 5e-5, beam size to 5, batch size to 8, training steps to 50K, and validation steps to 1K.

To have a very simple baseline, we developed an LSTM-based seq2seq model with OpenNMT-py \cite{klein2017opennmt} that follows the Sequencer's \cite{chen2019sequencer} bidirectional LSTM encoder, LSTM decoder, global attention, and copy selector, but does not use their abstraction phase and copy mechanism. The beam size was set to 5, batch size to 32, dropout to 0.3, training steps to 20K, and validation steps to 1K. The source code of this study is available at \cite{githubcode}.

We used the Accuracy metric to measure the performance of our model. Accuracy measures whether the generated fix codes are exactly the same as the actual fix codes implemented by the developers or not. There are other evaluation metrics like running test cases to find if the generated patches can pass the tests, but they are not always applicable since the projects should have a reliable test suite, and also there should be at least one test case to reflect the bug. Due to the lack of a complete test suite in the projects of datasets, we did not use testing-based evaluation metrics.

In addition, we compared the required input formats of our approach and some related work (e.g., SequenceR \cite{chen2019sequencer}, DLFix \cite{li2020dlfix}, and CoCoNut \cite{lutellier2020coconut}). Then, we discussed about the training and inference time of our approach, and the required additional efforts to support other programming languages. 

\subsection{Results}

\subsubsection{Answer to RQ1}

Table \ref{tab_evaluation_results} shows the result of our experiments on four different datasets considering top-1 predictions when accuracy is calculated. The results show that our approach works well on duplicate datasets (72\% and 68.8\%), but it has also an acceptable prediction accuracy on unique datasets (23.27\% and 19.65\%).

The accuracy for the simple baseline (2.2\%-17.65\%) is also shown in Table \ref{tab_evaluation_results}, which indicates the superiority of the CodeBERT approach. Especially, in cases where the dataset is relatively small, using a pre-trained model like CodeBERT benefits us more.

\begin{table}[htbp]
\caption{Evaluation Results, in terms of Accuracy}
\begin{center}
\begin{tabular}{|c|l|l|l|}
\hline
\textbf{Approach} & \multicolumn{2}{c|}{\textbf{Dataset}} & \multicolumn{1}{c|}{\textbf{Accuracy}} \\ \hline
\multirow{4}{*}{CodeBERT} & \multirow{2}{*}{Unique} & Large & 570 / 2449 (23.27\%) \\ \cline{3-4} 
 &  & Small & 92 / 468 (19.65\%) \\ \cline{2-4} 
 & \multirow{2}{*}{Duplicate} & Large & 4195 / 5820 (72\%) \\ \cline{3-4} 
 &  & Small & 569 / 827 (68.8\%) \\ \hline
\multirow{4}{*}{Simple Seq2Seq} & \multirow{2}{*}{Unique} & Large & 55 / 2449 (2.2\%) \\ \cline{3-4} 
 &  & Small & 16 / 468 (3.41\%) \\ \cline{2-4} 
 & \multirow{2}{*}{Duplicate} & Large &
 376 / 5820 (6.5\%) \\ \cline{3-4} 
 &  & Small & 146 / 827 (17.65\%) \\ \hline
\end{tabular}
\label{tab_evaluation_results}
\end{center}
\vspace*{-10pt}
\end{table}

As explained in the related work section, the proposed dataset is not readily usable with the previous works, so there is no easy way to compare our results with other baselines. Therefore, to put our results in the context we only look at the accuracy values reported in the related work (on their own datasets) to understand what is a typical range of accuracy achieved by a program repair tool.
Table \ref{tab_preivous_work_accuracy} shows the reported accuracy values of two tools on four datasets (two per method). The accuracy values are roughly in the same range (3.33\% to 27.33\%) as ours (19.65\% and 23.27\%) on the unique dataset (the duplicate dataset results are much higher). Though a direct comparison is not possible, still this gives us a context for further analysis. To compare our approach with alternatives, let's look at two recent machine learning-based approaches:

It takes 29 minutes for Sequencer \cite{chen2019sequencer} (see related work for more details) to generate the abstract buggy context of 75 bugs, which means that each bug needs 23 seconds, so if we want to use this method for our unique large dataset consisting of 2,449 bugs, it will roughly take more than 938 hours. Our technique does not require any abstraction process and it will take only 10 minutes to generate patches for the mentioned dataset, so our approach takes considerably less time.

Our technique can generate patches containing up to 510 tokens (CodeBERT has been trained with sequence lengths up to 512 tokens). Tufano et al. \cite{tufano2018empirical} address the unlimited vocabulary problem by renaming rare identifiers with a custom abstraction process, but since our technique uses a byte-level byte-pair-encoding (BBPE) as a tokenizer, it will progressively falls-back on character level embeddings for unseen words \cite{liu2019roberta}. This helps us to avoid any custom abstraction process or post-processing for removing unknown words.

We measured the execution time of our technique, running on a server with four GPUs (NVIDIA Tesla V100 SXM2 16GB). The execution time of training (5.5-9 hours) and inference (1.6-20 minutes) steps are shown in Table \ref{tab_execution_time}. We believe this is a reasonable time since fine-tuning the model is a one-time cost, but generating the patches, which is a repetitive task can be done very fast in less than a second, per bug.

\begin{table}[htbp]
\vspace*{-19pt}
\caption{The Accuracy Range of Previous Works}
\begin{center}
\resizebox{\linewidth}{!}{%
\begin{tabular}{|l|l|l|l|}
\hline
\multicolumn{1}{|c|}{\textbf{Name}} & \multicolumn{1}{c|}{\textbf{Beam Size}} & \multicolumn{1}{c|}{\textbf{Dataset}} & \multicolumn{1}{c|}{\textbf{Accuracy}} \\ \hline
\multirow{2}{*}{Sequencer \cite{chen2019sequencer}} & \multirow{2}{*}{50} & Defects4J & 18\% \\ \cline{3-4} 
 &  & CodRep4 & 20\% \\ \hline
\multirow{2}{*}{Tufano et al. \cite{tufano2018empirical}} & \multirow{2}{*}{5} & BFP & \begin{tabular}[c]{@{}l@{}}27.33\% (small) - 13.12\% (large)\end{tabular} \\ \cline{3-4} 
 &  & CodRep & \begin{tabular}[c]{@{}l@{}}10.27\% (small) - 3.33\% (large)\end{tabular} \\ \hline
\end{tabular}%
}
\label{tab_preivous_work_accuracy}
\end{center}
\end{table}

\begin{table}[htbp]
\caption{Training and Inference Execution Time}
\begin{center}
\begin{tabular}{|l|l|l|}
\hline
\multicolumn{1}{|c|}{\textbf{Dataset}} & \multicolumn{1}{c|}{\textbf{Training Time}} & \multicolumn{1}{c|}{\textbf{Inference Time}} \\ \hline
Duplicate - Large & 9 Hours & 20 Minutes \\ \hline
Unique - Large & 8 Hours & 10 Minutes \\ \hline
Duplicate - Small & 6.5 Hours & 2.5 Minutes \\ \hline
Unique - Small & 5.5 Hours & 1.6 Minutes \\ \hline
\end{tabular}
\label{tab_execution_time}
\end{center}
\vspace*{-5pt}
\end{table}

Since our technique does not require additional post-processing steps like removing unknown words or extra spaces, no more time is needed. Note that validating the generated patches with dynamic analysis like running test cases needs extra time, which is not part of our method's execution time.

\subsubsection{Answer to RQ2}

We found that our model can fix different types of bugs even if there are few instances of that bug type in the local training dataset. E.g., our technique fixes 60\% of the \verb#SWAP_BOOLEAN_LITERAL# bugs of the unique large dataset, but there are only 103 instances of this bug type in the local training dataset. This is due to the fact that CodeBERT has been trained with a large dataset, so having local training datasets with different types of bugs is not a necessity. Furthermore, while reviewing the names of the fixed bug projects, we came to the conclusion that the number of bug instances from one specific project in the local training dataset does not impact the effectiveness of our approach because it does not depend on the knowledge of the fixing patterns within a specific project. E.g., our model generated patches for 100\% of the bugs of \verb#Trinea.android-common# and \verb#alibaba.cobar# projects, but there is no instance from these projects in our local training dataset.

Also, our approach can generate codes that are not available in the local training dataset. This indicates its efficiency in generating completely new code. For example, our approach generates \verb#mReadOwners!=null && includingOwners# as the fixed code of  \verb#mReadOwners!=null#. The model fixes the if condition bug by adding a logical AND operator and an operand, but this line is not in our local training nor evaluation datasets. This patch is for \verb#android.platform_frameworks_base# project with \verb#29B2516012CF# (first 12 character) fixCommitSHA1 value.

Overall, our approach can generate 100\% accurate patches for different bug types, with a success rate in a range of 3.36\% to 77.63\% for unique datasets and 25\% to 94.12\% for the duplicate datasets. Table \ref{tab_top_3} shows the results of our top-3 fix categories. E.g.,  \verb#SWAP_BOOLEAN_LITERAL# indicates a bug where the True and False values should be exchanged. This is very popular bug in Java projects since Boolean flags are a code smell that reduces the code readability. 

\begin{table}[htbp]
\caption{Perfect Prediction Success Rate for Our Top-3 Bug Types}
\begin{center}
\resizebox{\linewidth}{!}{%
\begin{tabular}{|l|l|l|}
\hline
\multicolumn{1}{|c|}{\textbf{Dataset}} & \multicolumn{1}{c|}{\textbf{Bug Type}} & \multicolumn{1}{c|}{\textbf{Ratio}} \\ \hline
\multirow{3}{*}{Duplicate - Large}     & DIFFERENT\_METHOD\_SAME\_ARGS          & 92.35\%                             \\ \cline{2-3} 
                                       & CHANGE\_CALLER\_IN\_FUNCTION\_CALL     & 90.54\%                             \\ \cline{2-3} 
                                       & CHANGE\_OPERAND                        & 88.42\%                             \\ \hline
\multirow{3}{*}{Unique - Large}        & SWAP\_BOOLEAN\_LITERAL                 & 77.63\%                             \\ \cline{2-3} 
                                       & CHANGE\_OPERATOR                       & 46.15\%                             \\ \cline{2-3} 
                                       & DIFFERENT\_METHOD\_SAME\_ARGS          & 35.77\%                             \\ \hline
\multirow{3}{*}{Duplicate - Small}     & DIFFERENT\_METHOD\_SAME\_ARGS          & 94.12\%                             \\ \cline{2-3} 
                                       & CHANGE\_CALLER\_IN\_FUNCTION\_CALL     & 93.33\%                             \\ \cline{2-3} 
                                       & CHANGE\_OPERAND                        & 80\%                                \\ \hline
\multirow{3}{*}{Unique - Small}        & SWAP\_BOOLEAN\_LITERAL                 & 60\%                                \\ \cline{2-3} 
                                       & CHANGE\_OPERATOR                       & 31.82\%                             \\ \cline{2-3} 
                                       & SWAP\_ARGUMENTS                        & 30\%                                \\ \hline
\end{tabular}%
}\label{tab_top_3}
\end{center}
\vspace*{-10pt}
\end{table}

\begin{table}[htbp]
\caption{The Token Length of Generated Patches}
\begin{center}
\begin{tabular}{|l|l|l|}
\hline
\textbf{Project} & \textbf{FixCommitSHA1$^{\mathrm{a}}$} & \textbf{Length} \\ \hline
reactor.reactor-core & 20e155eeff37 & 193 \\ \hline
square.okhttp & f78f74f5a2cf& 192 \\ \hline
android.platform\_frameworks\_base & 946a17782a7a & 147 \\ \hline
clojure.clojure & 55ed50c4975c & 142 \\ \hline
oracle.graal & 2104049f33ef & 137 \\ \hline
\multicolumn{3}{l}{$^{\mathrm{a}}$Hash string contains the first 12 characters of the original value}
\end{tabular}
\label{tab_token_length}
\end{center}
\vspace*{-20pt}
\end{table}

Also, the \verb#SWAP_ARGUMENTS# indicates a bug where developers use a correct method but with the wrong order of arguments. This is again very common in Java projects since Java does not support named parameters, and it is the duty of developers to maintain parameters order. 

Finally, we analyzed the length of our generated fixes, and some of the results are shown in Table \ref{tab_token_length}. We found that our model can generate large fixed code as well as small code. As it is shown in Table \ref{tab_token_length}, our model can generate fixes with long lengths such as 193 for the bug in \verb#refactor.refactor-core#. For instance, we have got up to 50\% success rate (100\% accurate patches) for long patches (longer than 100 tokens) in the small-duplicate dataset.

\section{Conclusion and Future Work}
We propose an APR approach by using a pre-trained neural network model called CodeBERT for fixing Java simple bugs. We fine-tune our model on both small and large ManySStuBs4J datasets to find its capability to generate patches. We found that our approach is a viable solution for fixing bugs since it can generate fix codes for different types of bugs and its effectiveness and efficacy is comparable with state-of-the-art techniques. It generates fix codes in 19-72\% of the cases with different types of our datasets, which are exactly the same as the fix codes implemented by developers.

In addition, our approach does not suffer from limitations such as having a special token, short token length limitation, and unknown vocabulary problem, which makes it more practical. Also, our approach does not need any post-processing step, and it can be applied to other programming languages, supported by CodeBERT, without any extra effort.

Furthermore, we examined the characteristics of the fixed bugs and found that our approach can generate fixes with variable lengths. Finally, we observe that the number of instances of a specific bug type or the number of bugs from a specific project in the local training dataset does not have a negative impact on our approach effectiveness.

In future work, we intend to apply our approach to popular datasets such as Defects4j. Also, we aim to use other evaluation metrics, such as testing-based metrics, and more datasets in other programming languages, supported by our model.

\section*{References}
\printbibliography[heading=none]

\end{document}